\documentclass[aps,pra,reprint,superscriptaddress]{revtex4-2}

\usepackage{color}
\usepackage{amsmath}
\usepackage{amssymb}
\usepackage{graphicx}
\usepackage[a4paper,top=2.00cm,bottom=2.00cm,left=2.00cm,right=2.00cm]{geometry}

\usepackage{color}
\usepackage{array}
\usepackage{multirow}
\usepackage{xspace}% correct spacing
\usepackage{bm}% bold math
\usepackage{xcolor}

\begin{document}
\title{Niobium in clean limit: an intrinsic type-I superconductor}
\author{Ruslan Prozorov}
\email{prozorov@ameslab.gov}

\affiliation{Ames National Laboratory, Ames, IA 50011, USA}
\affiliation{Department of Physics \& Astronomy, Iowa State University, Ames, IA
50011, USA}
\author{Mehdi Zarea}
\affiliation{Center for Applied Physics and Superconducting Technologies Department
of Physics and Astronomy Northwestern University, Evanston, IL 60208,
USA}
\author{James A. Sauls}
\affiliation{Center for Applied Physics and Superconducting Technologies Department
of Physics and Astronomy Northwestern University, Evanston, IL 60208,
USA}

\date{24 July 2022}

\begin{abstract}
Niobium is one of the most researched superconductors, both theoretically
and experimentally. It is enormously significant in all branches of
superconducting applications, from powerful magnets to quantum computing.
It is, therefore, imperative to understand its fundamental properties
in great detail. Here we use the results of recent microscopic calculations
of anisotropic electronic, phonon, and superconducting properties,
and apply thermodynamic criterion for the type of superconductivity,
more accurate and straightforward than a conventional Ginzburg-Landau
parameter $\kappa$ - based delineation, to show that pure niobium
is a type-I superconductor in the clean limit. However, disorder
(impurities, defects, strain, stress) pushes it to become a
type-II superconductor.
\end{abstract}
\maketitle

\section{Introduction}

Niobium metal is one of the most important materials for superconducting
technologies, from SRF cavities \cite{Gurevich2012}, to superconducting
circuits for sensitive sensing \cite{Clarke2004} and quantum informatics
\cite{Reagor2016}. Numerous experimental works report measurements
on different samples, from almost perfect single crystals to disordered
films \cite{Finnemore1966,Rollins1977,Rusnaka,Bahte1998,Reagor2016,Kozhevnikov2017,Koethe2000,Lechner2020,Lee2021,Ooi2021}.
Likewise, numerous theories explore its properties from microscopic
calculations to phenomenological theories \cite{Scott1970,Ohta1978,Butler1980,Daams1980,Gurevich2012,Liarte_2017,Kubo_2021,Zarea2022}.
It is impossible to acknowledge a multitude of relevant references,
so we will limit ourselves to the specific topic of the paper.

Despite various attempts, first-principle calculations of the absolute
values of the critical fields, in particular the upper critical field,
$H_{c2}$, remain in poor agreement with the experiment. As a result, either
the temperature dependence of the normalized field, usually as introduced
by Helfand and Werthamer, $h^{*}\equiv H_{c2}\left(T\right)/T_{c}H'_{c2}\left(T=T_{c}\right)$
\cite{HW1966}, is calculated \cite{Arai2004}, or calculations use
experimental parameters, such as Fermi velocities, $v$, to fit the
data \cite{Butler1980}. Considering that $H_{c2}\sim v^{-2}$, this
makes a significant difference. As we show in this paper, this is
no fault of the theorists, but rather quite ambiguous experimental
determination of the critical fields. This, in turn, is no fault of
the experimentalists, because it is clear that Nb is extraordinarily
susceptible to the disorder with its experimental $RRR\equiv R\left(300\:\text{K}\right)/R\left(T_{c}\right)$,
ranging between $3$ and $90000$ \cite{Finnemore1966,Koethe2000}.

Previously, the problem of the identification of the type of superconductivity
was analyzed in detail at arbitrary temperatures \cite{KoganCriterion2014,KP2014}.
It was argued that instead of a conventional criterion based on the
Ginzburg-Landau parameter, $\kappa=\lambda/\xi$, one has to use the
ratio of the upper and thermodynamic critical fields, $h_{c2,c}\equiv H_{c2}/H_{c}$.
Alternatively, it could be the ratio of $h_{c,c1}\equiv H_{c}/H_{c1}$,
but superheating \cite{PhysRevB.53.5650,MATRICON1967241} and various
surface barriers \cite{Bean1964a,Brandt1995} make experimental determination
of the lower critical field, $H_{c1}$, difficult. Only when $h_{c2,c}>1$,
do vortices form and the material can be identified as a type-II superconductor.
As it is shown in Ref. \cite{KoganCriterion2014,KP2014}, the $\kappa-$based
criterion coincides with the thermodynamic criterion only at $T_{c}$,
and not even within a small temperature interval below $T_{c}$. In
other words, the slopes of the temperature-dependent $h_{c2,c}\left(T\right)$
and $\kappa\left(T\right)$ are different at $T_{c}$. In anisotropic
superconductors, the same sample can be type-I in one orientation
of the magnetic field, and type-II in another \cite{KoganCriterion2014}.

Of course, in the case of significantly different $\lambda$ and $\xi$,
the difference between the two criteria is not that important and
this is why the type of most superconductors was correctly identified using the Ginzburg-Landau parameter $\kappa$. Moreover, it is well known that there are practically no proven non-elemental type-I
superconductors, except for a few suggested compounds, such as Ag$_{5}$Pb$_{2}$O$_{6}$
\cite{Yonezawa2005}, YbSb$_{2}$ \cite{Zhao2012}, OsB$_{2}$ \cite{Bekaert2016},
and PdTe$_{2}$ \cite{Leng2017}. However, until the intermediate state is directly observed instead of a mixed state of Abrikosov vortices by magneto-sensitive imaging techniques, the type-I status of these materials will remain ``pending''.

Niobium seems to be a difficult case, because, by all accounts, it
is situated close to the crossover boundary and can be easily moved
deeper into the type-II side by non-magnetic disorder, which increases
the London penetration depth, $\lambda$, and decreases the coherence
length, $\xi$. Magnetic disorder, on the other hand, pushes a superconductor
into the opposite direction Recently high-quality magneto-optical
imaging of Nb single crystals with $RRR=500$ has revealed directly
and unambiguously a clear structure of the intermediate state \cite{Ooi2021}.
These images are strikingly similar to images by one of us (RP) for the commonly
accepted type-I superconductor, pure lead \cite{Prozorov2007,Prozorov2008a,Prozorov2005}.
So similar that the authors of Ref.\cite{Ooi2021} write in their
paper, ``\emph{The observed patterns of the IMS {[}intermediate mixed
state{]} are rather similar to that of the Meissner and normal domains
in the intermediate state of the type-I superconductor, Pb reported
by Prozorov et al.} \cite{Prozorov2007,Prozorov2008a}''. This is,
indeed, the case.

The tendency to type-I behavior of elemental metals is not unique to niobium. A clear cross-over from a type-I (confirmed by magneto-optical imaging \cite{Essmann1977}) to a type-II regime upon introduction of non-magnetic scattering was convincingly demonstrated in tantalum \cite{Essmann1977,Idczak2020}. In known type-II vanadium, the Ginzburg-Landau parameter $\kappa$ approaches the borderline value of $1/\sqrt{2}$ toward type-I behavior with the increase of the residual resistivity ratio \cite{Weber1981}.

\begin{figure}[tbh]
\centering \includegraphics[width=8.5cm]{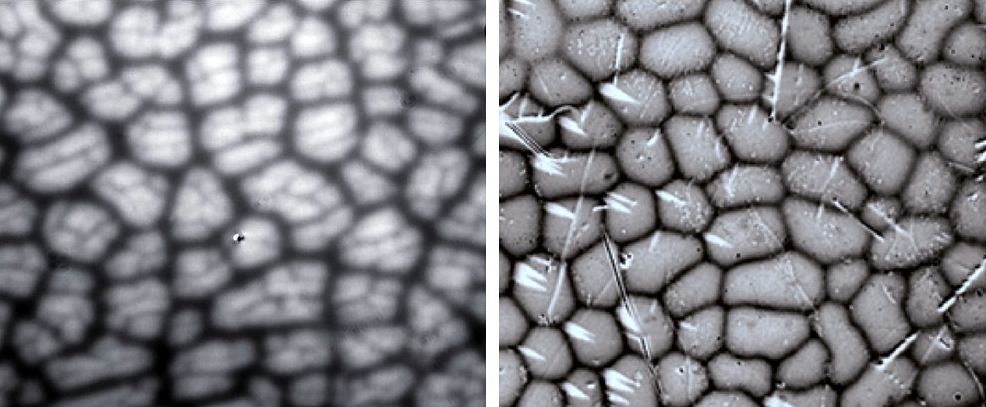} \caption{(Left) the intermediate state in a single crystal Nb (Ref.\cite{Ooi2021},
Fig.3a). (Right) the intermediate state in a single crystal Pb, (Ref.\cite{Prozorov2008a},
Fig.2g). {[}Left frame reprinted with permission from: S. Ooi, M.
Tachiki, T. Konomi, T. Kubo, A. Kikuchi, S. Arisawa, H. Ito, and K.
Umemori, ``\emph{Observation of Intermediate Mixed State in High-Purity
Cavity-Grade Nb by Magneto-Optical Imaging}'', Physical Review B
\textbf{104}, 6 (2021). Copyright (2021) by the American Physical
Society{]}}
\label{fig1}
\end{figure}

Figure \ref{fig1} compares the intermediate state structure in niobium
and lead single crystals. The visual similarity is remarkable. Of
course, depending on the material, its properties and the proximity
to the crossover boundary the fine details vary. For example, here
the field-cooled (FC) image is shown for Nb, and zero-field-cooled
is shown for Pb. Upon field cooling, the intermediate state in Pb breaks into a corrugated
laminar structure, whereas in Nb it apparently further breaks into
large flux tubes, only reinforcing our earlier conclusion that the
equilibrium topology of the intermediate state in type-I superconductors
is tubular, rather than laminar \cite{Prozorov2007}. Since the textbooks
tell us that the true intermediate state structure is described as laminar, stripy, or labyrinth-like, the observation of the tubular features was often
interpreted as some kind of crossover ``intermediate mixed state''
when vortices gather into ``domains'' as illustrated in Fig.4e of
Ref.\cite{Ooi2021}. However, no evidence of individual vortices was
found in such structures. For a variety of magneto-optical images
of superconductors, the reader is referred to Ref. \cite{Huebener2001}.
In our interpretation, the authors of Ref.\cite{Ooi2021} have observed
a genuine intermediate state in a clean-limit Nb crystal proving
experimentally that Nb is a type-I superconductor. We now check whether
the microscopic theory agrees.

\section{Properties of niobium from the microscopic theory}

Recently, based on the density functional theory (DFT) calculations
of the electron and phonon band structures, microscopic superconducting
properties of elemental niobium were determined using Eliashberg formalism
\cite{Zarea2022}. It was found that pure Nb is a two-active-bands,
two-gap superconductor. The bands are moderately anisotropic with
temperature-dependent anisotropies. The more isotropic band 2 dominates
the electronic properties. For analytical estimates, we use isotropic
BCS formulas, but then we use 2-band averaged RMS values from Ref.
\cite{Zarea2022}. Specifically, $v=\sqrt{\left\langle v^{2}\right\rangle }$,
the RMS value of the Fermi velocity is given by, $v=\sqrt{n_{1}v_{1}^{2}+n_{2}v_{2}^{2}}$,
where partial densities of states (DOS), $n_{1,2}=N_{1,2}(0)/\left(N_{1}+N_{2}\right)$,
and a similar equation was used for the RMS superconducting gap. We
compare analytical results with the full numeric evaluation of anisotropic
equations. The paper uses cgs units throughout. We calculate critical
fields analytically at $T=0$ and $T=T_{c}$ and their ratios and
compare them with the Ginzburg-Landau criterion for the type of a superconductor.

Here we summarize the parameters used from Ref.\cite{Zarea2022}.
The densities of states at the Fermi level: $N_{1}(0)=6.33$$\times10^{33}$
erg$^{-1}$cm$^{-3}$, $N_{2}(0)=3.98$$\times10^{34}$ erg$^{-1}$cm$^{-3}$,
$N_{tot}(0)=4.61$$\times10^{34}$ erg$^{-1}$cm$^{-3}$. The averaged
RMS velocities, $,v_{1}=4.37$$\times10^{7}$ cm/s, $v_{2}=7.62$$\times10^{7}$
cm/s, and 2-band average, $v=7.26\times10^{7}cm/s$. The RMS averaged
superconducting gaps, $\varDelta_{1}\left(0\right)=3.14\times10^{-15}$
erg, $\varDelta_{2}\left(0\right)=2.53\times10^{-15}$ erg, and $\varDelta\left(0\right)=2.62\times10^{-15}$
erg . For comparison, the weak-coupling BCS gap is $\varDelta_{BCS}\left(0\right)=1.7638T_{c}=2.27\times10^{-15}$
erg with $T_{c}=9.33$ K. The Fermi energy of Nb is $E_{F}=5.32$
eV $=8.52\times10^{-12}$ erg \cite{HCP}. The total carrier density
is $n=5.56\times10^{22}$ cm$^{-3}$ and the effective electron mass
is $m^{*}=2.14m_{e}$ \cite{Scott1970}. Note that two-bands averaged
values are very close to band 2 values, reflecting its dominant character.

Let us now estimate various quantities using analytic limiting cases
from the BCS theory. The upper critical field at $T=0$ is given by \cite{HW1966},

\[
\ensuremath{H_{c2}\left(0\right)=\frac{\phi_{0}}{2\pi\xi^{2}}=\frac{\phi_{0}\pi k_{B}^{2}T_{c}^{2}}{2\hbar^{2}v^{2}}\exp\left(2-C\right)}
\]
where $C=0.577216$ is the Euler constant. Technically, two bands
will have two different characteristic coherence lengths \cite{Gurevich2003}.
Of course, there is only one upper critical field, but a formal substitution
of bands' Fermi velocities gives $H_{c2}\left(0\right)=1053$ Oe
and $346$ Oe for bands 1 and 2, respectively. These values are far from the
reported values of $3-5$ kOe \cite{Arai2004,Butler1980,Finnemore1966}. For
the 2-band average, $H_{c2}\left(0\right)=381$ Oe. The coherence
lengths formally corresponding to these fields, are $\xi\left(0\right)=56$
nm and $97$ nm for bands 1 and 2, respectively, and $\xi\left(0\right)=93$
nm for the two-band average. For comparison, the BCS coherence length
is:

\[
\ensuremath{\xi_{0}=\frac{\hbar v}{\pi\varDelta\left(0\right)}}
\]
which gives $\xi_{0}=47$ nm and $101$ nm for bands 1 and 2, respectively,
and $\xi_{0}=98$ nm for the 2-band average.

The thermodynamic critical field is given by

\[
\ensuremath{H_{c}\left(0\right)=2\sqrt{\pi N\left(0\right)}\sqrt{\left\langle \varDelta^{2}\left(0\right)\right\rangle }}
\]
and we obtain $H_{c}\left(0\right)=1993$ Oe. Note a substantial
difference between this value and the much lower ``upper'' critical
fields above. On the other hand, this field scale is quite close to
what was determined as experimental critical fields in clean samples
\cite{Finnemore1966,Ohta1978}, considering that it is very difficult
to distinguish hysteresis loops of relatively pure niobium and, for
example, lead with some pinning \cite{Prozorov2007}.

\section{Type-I superconductivity in clean-limit niobium metal}

According to Ref.\cite{KoganCriterion2014}, the natural way to determine
the type of a superconductor is to examine the ratio of $h_{c2,c}\equiv H_{c2}/H_{c}$. While Ref.\cite{KoganCriterion2014}, provides a recipe
for calculating this ratio at all temperatures, here we only need
to consider $T=0$ and $T=T_{c}$ for which analytic expressions are
available. If the gap anisotropy is described by the order parameter,
$\varDelta\left(T,k\right)=\Psi\left(T\right)\Omega\left(k\right)$,
where the angular part is normalized over Fermi surface average, $\left\langle \Omega^{2}\right\rangle =1$,
then for the magnetic field along the $c-$axis \cite{KoganCriterion2014},

\begin{equation}\label{eq-hc2c0}
\ensuremath{h_{c2,c}\left(0\right)=\frac{\phi_{0}k_{B}T_{c}}{\hbar^{2}v_{0}^{2}\sqrt{\pi N\left(0\right)}}\exp\left\langle \Omega^{2}\ln\frac{\left|\Omega\right|}{\mu_{z}}\right\rangle }
\end{equation}
where $\mu_{z}=\left(v_{x}^{2}+v_{y}^{2}\right)/v_{0}^{2}$, and $v_{0}$
is the characteristic velocity scale (equal to Fermi velocity in isotropic
case),
\[
\ensuremath{v_{0}=\left(\frac{2E_{F}^{2}}{\pi^{2}\hbar^{3}N\left(0\right)}\right)^{1/3}}
\]

For band 2, we have: $v_{0,2}=6.81$$\times10^{7}$ cm/s and using
total DOS, $v_{0,tot}=6.48$$\times10^{7}$ cm/s. In particular, for
the isotropic case, $\left\langle \mu_{c}\right\rangle =2/3$, and
$\left\langle \ln\mu_{c}\right\rangle =2\left(\ln2-1\right)$, which
then reproduces Helfand-Werthamer result \cite{HW1966},

\[
\ensuremath{h_{c2,c}\left(0\right)=\frac{\phi_{0}k_{B}T_{c}}{\hbar^{2}v_{0}^{2}\sqrt{\pi N\left(0\right)}}\exp\left(-2\left(\ln2-1\right)\right)}
\]
Using the two-band average, we obtain, $h_{c2,c}^{2bands}\left(0\right)=0.221$, while
using only band 2 we obtain $h_{c2,c}^{band2}\left(0\right)=0.216$.

For $T\rightarrow T_{c}$ we have in general:
\begin{equation}
\ensuremath{h_{c2,c}\left(T_{c}\right)=\frac{3\sqrt{2}\phi_{0}k_{B}T_{c}}{\hbar^{2}v_{0}^{2}\sqrt{7\zeta\left(3\right)\pi N\left(0\right)}}\frac{\sqrt{\left\langle \Omega^{4}\right\rangle }}{\left\langle \Omega^{2}\mu_{z}\right\rangle }}
\,
\end{equation}
where $\zeta\left(3\right)=1.2021$ is Riemann's zeta function. In
the isotropic case this reduces to,
\[
\ensuremath{h_{c2,c}\left(T_{c}\right)=\frac{3\sqrt{2}\phi_{0}k_{B}T_{c}}{\hbar^{2}v_{0}^{2}\sqrt{7\zeta\left(3\right)\pi N\left(0\right)}}}
\]
For the two-band average we obtain $h_{c2,c}^{2bands}\left(T_{c}\right)=0.175$, and for band 2,
$h_{c2,c}^{band2}\left(T_{c}\right)=0.171$. Looking at the previous
two equations it is easy to see that their ratio is a pure number,
$h_{c2,c}\left(0\right)/h_{c2,c}\left(T_{c}\right)=1.263$, independent
of material properties. However, the ratio does depend on the anisotropy
of the Fermi surface and of the order parameter via Fermi surface
averages of the terms containing functions of $\Omega(k)$.
Indeed the result for the Fermi surface average appearing in Eq.~\eqref{eq-hc2c0} based on the bandstructure and Eliashberg results for the Fermi velocity and anisotropic gap function at $T/T_c=0.32$ yield,
\begin{equation}
    \left\langle\Omega^2\,\ln\left(\frac{|\Omega|}{\mu_z}\right)\right\rangle = 0.98
    \,,
\end{equation}
which gives $h_{c2,c}(0)\approx 0.319$. Thus, pure, single-crystaline Nb is in the Type I limit based first-principles calculations of the gap and Fermi surface anisotropy~\cite{Zarea2022}.

An additional method to  verify the consistency of the above analysis is to use the
fact that at $T=T_{c}$ \cite{KoganCriterion2014},
\[
h_{c2,c}\left(T_{c}\right)=\sqrt{2}\kappa_{GL}
\]
where Ginzburg-Landau $\kappa_{GL}$ is given by,
\[
\ensuremath{\kappa_{GL}=\frac{3\phi_{0}k_{B}T_{c}}{\hbar^{2}v^{2}\sqrt{7\zeta\left(3\right)\pi N\left(0\right)}}}
\]
Evaluating for two bands average, we obtain at $T_{c}$, $\kappa_{GL}=0.123$
and then, indeed $\sqrt{2}\kappa_{GL}=0.175=h_{c2,c}\left(T_{c}\right)$.
In another limit, $T=0$, we can evaluate $\kappa\left(0\right)=\lambda\left(0\right)/\xi\left(0\right)$.
In the isotropic approximation, the London penetration depth becomes,
\[
\lambda\left(0\right)=\frac{c}{e}\sqrt{\frac{m^{*}}{4\pi n}}\approx33\:\mathrm{nm}
\]
Thus,
\[
\kappa\left(0\right)=\frac{33\:\mathrm{nm}}{93\:\mathrm{nm}}=0.355<\frac{1}{\sqrt{2}}=0.71
\]
Therefore, even this simple estimate based on isotropic London theory
gives the value of $\kappa$ corresponding to type-I superconductivity.
A more straightforward thermodynamic criterion based on the ratio of the critical fields
gives the values of $h_{c2,c}$ clearly lower than one, which places
niobium in the domain of type-I superconductivity. The situation, however,
quickly changes with the addition of non-magnetic scattering. The upper critical
field grows linearly with the scattering rate \cite{Gurevich2003,Kogan2013,Xie2017}
and quickly exceeds $H_{c}$. Magnetic impurities, on the other hand,
would bring $H_{c2}$ down, but they will also suppress
$T_{c}$ \cite{Kogan2013,KP2014}.

For applications of SRF cavities for particle accelerator technology it
is desirable to stabilize Nb closer to a type-I phase where one can
increase the superheating field by engineering the disorder profile very close to the superconducting-vacuum interface, leaving much of the London penetration region in the clean limit~\cite{nga19,gur17}. In quantum informatics
where thin films are used, perhaps switching to the epitaxial growth instead
of ablation-type sputtering would improve the RRR, and hence improved device performance.

\section{Conclusions}

It is shown that using the parameters of recent microscopic calculations
of superconducting and electronic properties of pure Nb, the estimated
ratio of the upper and thermodynamic critical fields, $H_{c2}/H_{c}$,
changes from 0.22 at $T=0$, to 0.18 at $T_{c}$. These values place
clean-limit niobium squarely into the domain of type-I superconductivity.
This conclusion is firmly supported by the direct magneto-optical observation
of the intermediate state in Nb single crystals with RRR=500 \cite{Ooi2021}.
It is suggested that ever-present disorder and impurities drive the
real material to a type-II side in most samples studied by experimentalists so far.

\begin{acknowledgments}

We thank Vladimir Kogan and Alex Gurevich for useful discussions.
This work was supported by the U.S. Department of Energy, Office of
Science, National Quantum Information Science Research Centers, Superconducting
Quantum Materials and Systems Center (SQMS) under contract number
DE-AC02-07CH11359. Ames Laboratory is operated for the U.S. DOE by Iowa State University under contract \# DE-AC02-07CH11358.
\end{acknowledgments}

%\bibliographystyle{apsrev4-2}
%\bibliography{Nb-SQMS}


\end{document}